\documentclass[twocolumn,amsmath,amssymb,footinbib,superscriptaddress]{revtex4-2}

\usepackage{graphicx,color}
\usepackage{dcolumn}
\usepackage{bm}

\usepackage{epstopdf}
\usepackage{amsmath}
\usepackage[colorlinks=true,citecolor=blue,urlcolor=blue,linkcolor=blue]{hyperref}

\begin{document}

\title{Strongly coupled interface electronic states and interface phonon mode at GaP/Si(001)}

\author{Gerson Mette}
\affiliation{Faculty of Physics and Materials Sciences Center, Philipps-Universit{\"a}t Marburg, 35032 Marburg, Germany}

\author{Kunie Ishioka}
\affiliation{Nano-characterization Unit, National Institute for Materials Science, Tsukuba, 305-0047 Japan}

\author{Steven Youngkin}
\affiliation{Faculty of Physics and Materials Sciences Center, Philipps-Universit{\"a}t Marburg, 35032 Marburg, Germany}


\author{Wolfgang Stolz}
\affiliation{Faculty of Physics and Materials Sciences Center, Philipps-Universit{\"a}t Marburg, 35032 Marburg, Germany}

\author{Kerstin Volz}
\affiliation{Faculty of Physics and Materials Sciences Center, Philipps-Universit{\"a}t Marburg, 35032 Marburg, Germany}

\author{Ulrich H{\"o}fer}
\affiliation{Faculty of Physics and Materials Sciences Center, Philipps-Universit{\"a}t Marburg, 35032 Marburg, Germany}

\date{\today}

\begin{abstract}

Ultrafast carrier and phonon dynamics at the buried
heterointerface of GaP/Si(001) are investigated by means of
two-color pump-probe reflectivity measurements. The
carrier-induced reflectivity signal exhibits a resonant
enhancement at pump-photon energies of 1.4~eV, which can be
assigned to an optical transition between electronic interface
states. The transient reflectivity is modulated by a coherent
oscillation at 2~THz, whose amplitude also becomes maximum at
1.4~eV. The observed resonant behavior of the phonon mode in
combination with a characteristic wavelength-dependence of, both,
its frequency and initial phase, strongly indicate that the 2-THz
mode is a difference-combination mode of a GaP-like and a Si-like
phonon at the heterointerface and that this second-order
scattering process can be enhanced by a double resonance involving
the interfacial electronic states.

\end{abstract}

\maketitle


Coupling of charge and lattice degrees of freedom in semiconductors
is one of the key factors to determine their crystalline structure
and dominate the electronic, optical and thermal
properties~\cite{YuCardona}. At surfaces of inorganic
semiconductors, it has been established that the rearrangement of
the atomic structure leads to well-defined electronic and phononic
surface states~\cite{Monch}. In comparison, deeply buried interfaces
have posed greater challenges due to the limitation of applicable
experimental techniques and the complexity in theoretical
simulations. Heterointerfaces between non-polar (e.g. Si) and polar
(e.g. group III/V) semiconductors are particularly demanding,
because of the possible localization of electric charges at the
interface, the occurrence of anti-phase boundaries (APB), and
because of the lattice mismatch between two semiconductors with
different crystalline structures.

The interface of GaP and Si has recently gained attention because
the two semiconductors have small mismatch in their lattice
constants and can therefore serve as a template for Si-based III/V
optoelectronics~\cite{Beyer2019, Supplie2018}. To specify the atomic
bonds and electric charges at the heterointerface, the growth of a
pseudomorphic GaP nucleation layer on Si(001) was monitored
\textit{in situ} during the first growth steps by means of
low-energy electron diffraction and reflection anisotropy
spectroscopy, as well as simulated by density functional theory
calculations~\cite{Supplie2015, Romanyuk2016, Supplie2018}. The
atomic structures of the interface and the APB were determined by
transmission electron microscopy~\cite{Beyer2016, Beyer2019} and
cross-sectional scanning tunneling microscopy~\cite{Lenz2019,
Saive2018}. Direct access to the \emph{electronic} states at the
buried interface has been limited, however, due to the small escape
depth of photoelectrons in standard photoemission spectroscopy.

Nonlinear spectroscopic techniques such as second-harmonic
generation (SHG) have advantages in their accessibility and
sensitivity to buried interfaces.  A previous time-resolved SHG
study on a thin GaP nucleation layer on Si(001) revealed short-lived
($\lesssim400$ fs) electronic states that were resonantly excited at
pump-photon energies of 1.4\,eV~\cite{Mette2020}. This resonance was
attributed to electronic states which are spatially localized at the
heterointerface and which lie energetically in the band gap of the
two semiconductors. This previous study observed no apparent lattice
vibration, however, in spite of a general expectation for the
emergence of interfacial phonon modes.

Transient reflectivity has been extensively applied to study carrier
and phonon dynamics of semiconductor heterostructures \cite{Dekorsy,
Foerst}, though as a linear spectroscopic technique it has no
specific sensitivity to interfaces or surfaces. In the present
study, we investigate the interfacial carrier and phonon dynamics at
GaP/Si(001) by performing transient reflectivity measurements using
tunable near-infrared pump light.
The carrier-induced reflectivity response from a thin GaP nucleation
layer on Si(001) exhibits a clear resonance at 1.4\,eV in agreement
with the previous SHG measurements~\cite{Mette2020}. Moreover, the
transient reflectivity signals are modulated by an oscillation at
2\,THz, whose amplitude follows the same resonance as the
carrier-induced response. The consistent resonant behavior
unambiguously proves that this oscillation is an interface phonon
mode. The observed wavelength-dependence of, both, frequency and
initial phase of the 2-THz mode furthermore suggests that the
interfacial phonon is enhanced by a double resonance involving the
interface electronic states.


The studied sample is a nominally undoped 10-nm thick GaP nucleation
layer grown at 450$^\circ$C on $n$-type Si(001) by using flow-rate
modulated metal-organic vapor phase epitaxy (MOVPE)~\cite{Volz2011,
Beyer2012}. The GaP nucleation layer consists of crystalline grains
with a lateral size of $\leq$30\,nm as shown in Fig.~S1 in the
Supplementary Material (SM). Two-color pump-probe reflectivity
measurements are performed under ambient conditions in the near
back-reflection geometry. Two different sets of tunable femtosecond
light sources are employed to cover the pump wavelength range of
$720-1220$\,nm (photon energy of $1.02-1.72$\,eV), whose details are
described in Section~II of the SM. The pump-photon energies are
smaller than the indirect band gap of GaP (2.26\,eV), but comparable
with or larger than that of Si (1.12\,eV). The probe wavelength was
fixed at 800\,nm (1.55\,eV). The pump-induced change in the
reflectivity $\Delta R/R$ is measured by detecting the probe lights
before and after the reflection by the sample with a pair of matched
photodiodes. The time delay $t$ between the pump and probe pulses is
scanned using a linear motor stage with slow scan technique.


\begin{figure}
\includegraphics[width=0.475\textwidth]{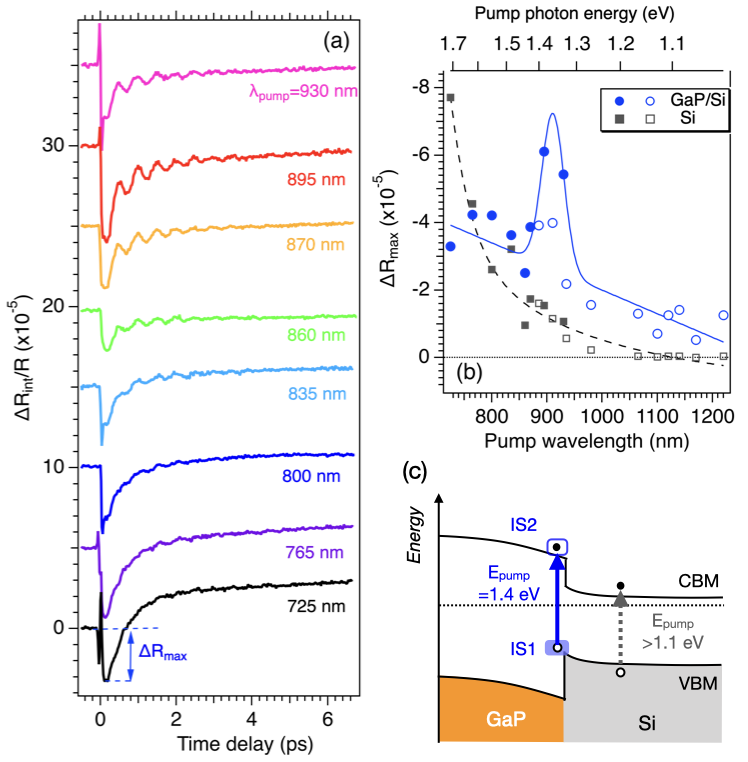}
\caption{\label{TDGaPSi_Si}
(a) Interface-contribution in transient reflectivity of the GaP/Si heterostructure
pumped at different wavelength and probed at 800\,nm. Traces are offset for clarity.
(b) Height of the initial drop $R_\text{max}$ of the interface contribution of GaP/Si
 as a function of the pump wavelength in comparison to bulk Si. Filled and open symbols
 are obtained with two different light sources. The latter are multiplied by 1.7 to
 scale to the former. Curves are to guide the eye. (c) Schematic band diagram of the
 GaP/Si interface. IS1 and IS2 indicate occupied and unoccupied interface states, respectively.}
\end{figure}

The as-measured transient reflectivity signals $\Delta R$ from the
GaP/Si sample (cf. Fig.~S2(a) in the SM) contain contributions not
only from the GaP overlayer/heterointerface $\Delta R_\text{int}$
but also from the Si substrate $\Delta R_\text{Si}$. Therefore, we
extract the former contribution by subtracting the latter from the
as-measured signal ($\Delta R_\text{int}\equiv\Delta R-\Delta
R_\text{Si}$) as described in more detail in Section~III of the SM.

Fig.~\ref{TDGaPSi_Si}(a) compares the obtained
overlayer/interface-contribution $\Delta R_\text{int}/R$ for
different pump wavelengths. The transients show an abrupt drop at
$t=0$ which is followed by a double exponential increase. The height
of the initial drop $\Delta R_\textrm{max}$, which gives a
semi-quantitative measure for the photoexcited carrier density in
the overlayer and at the heterointerface, exhibits a distinct
resonance peak at $\lambda_\textrm{pump}\simeq$900\,nm (photon
energy of 1.4\,eV) on top of the monotonic increase with decreasing
wavelength (increasing pump-photon energy) as plotted in
Fig.~\ref{TDGaPSi_Si}(b).  As shown in the same figure, this
resonance behavior is in apparent contrast to the initial drop
height of bulk Si (c.f. Fig.~S2(b) in SM for the corresponding
transients of bulk Si). It coincides with the resonance of the fast
SHG component obtained in our previous study on a thin ($d=4.5$\,nm)
GaP film on Si(001)~\cite{Mette2020}. We therefore attribute the
1.4-eV resonance to an optical transition between interface
electronic states as schematically shown in
Fig.~\ref{TDGaPSi_Si}(c). After the initial abrupt drop, $\Delta
R_\text{int}/R$ can be fitted to a multiple exponential function.
The time constant for the fast rise is $\sim0.2$\,ps, which is
comparable to the fast decay of the SHG signal observed
previously~\cite{Mette2020}. The slower rise occurs on a similar
time scale to that of bulk Si (cf. Fig.~S2(b) in SM), $\sim$40\,ps,
and is therefore attributed to the carriers photoexcited in the Si
substrate and recombining at the GaP/Si
interface~\cite{Sabbah2000,Sabbah2002,Ishioka2021}.

\begin{figure*}
\includegraphics[width=0.575\textwidth]{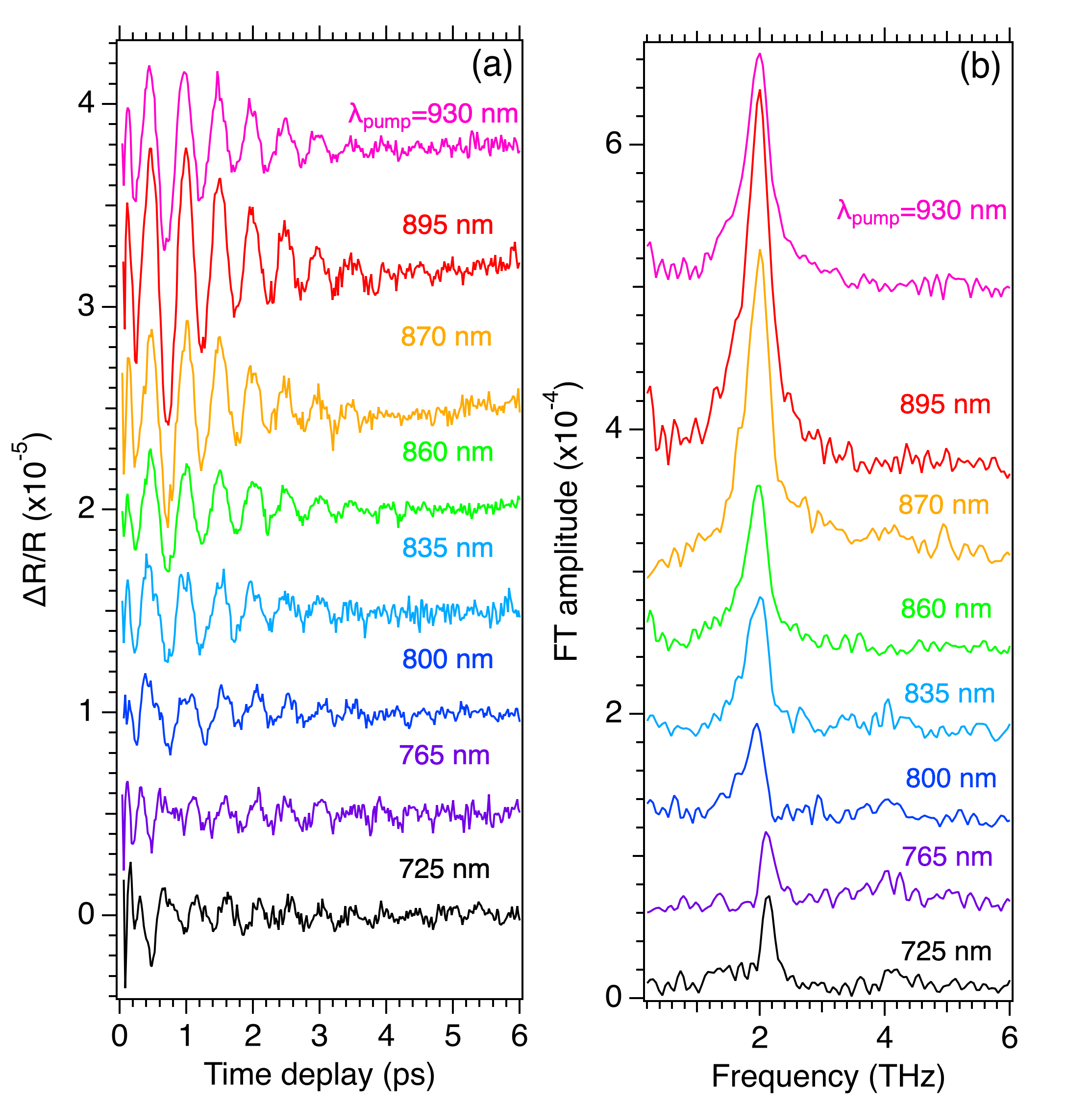}
\includegraphics[width=0.375\textwidth]{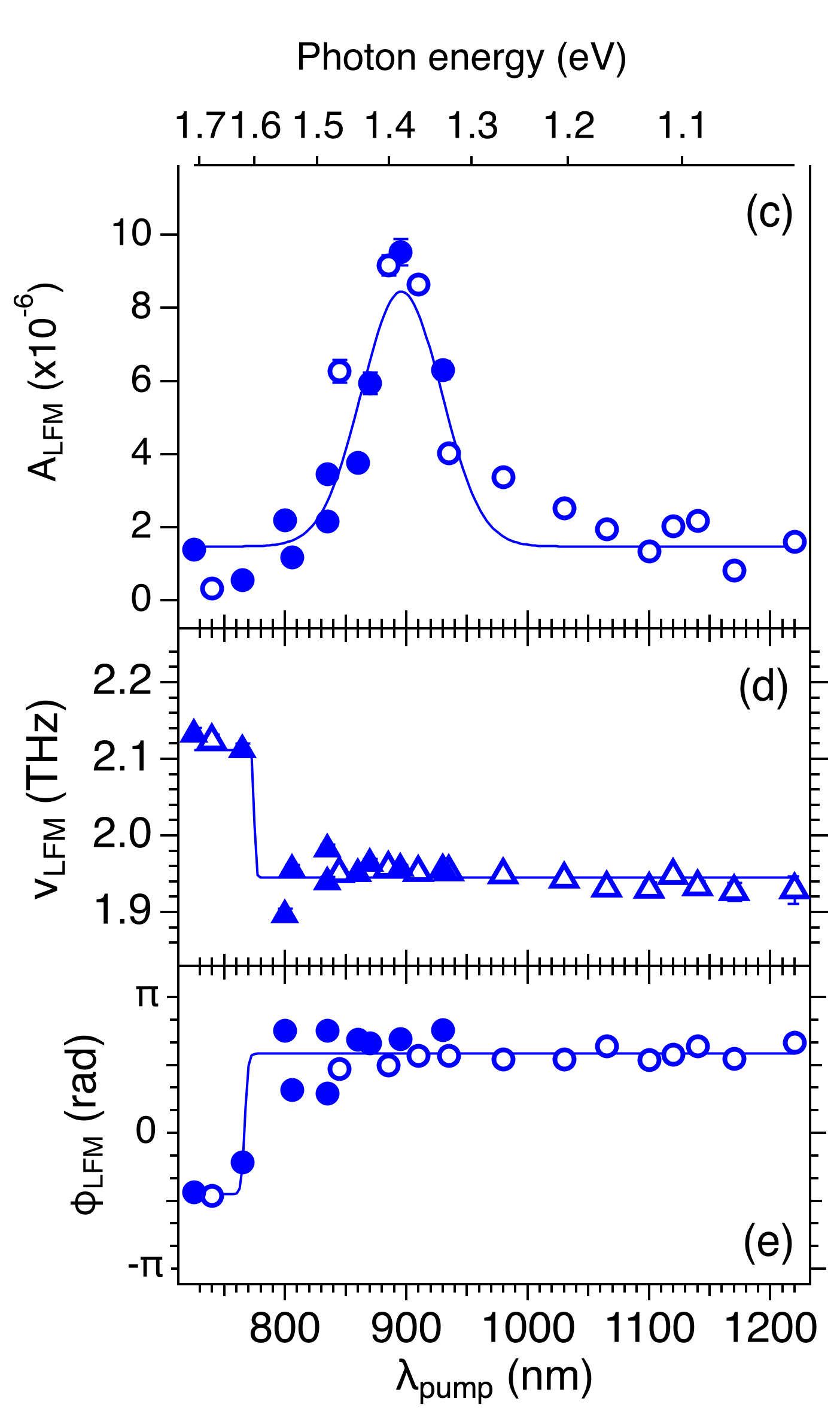}
\caption{\label{CFT}
(a) Oscillatory part of transient reflectivity of GaP/Si pumped at
different wavelengths and (b) the corresponding fast Fourier transform (FFT) spectra.
Traces are offset for clarity.
(c) Amplitude, (d) frequency, and (e) initial phase of the low-frequency mode (LFM) as a
function of the pump wavelength. Filled and open symbols in each panel represent
the results obtained with two different light sources. In (c) the latter are
multiplied by 1.7 to scale to the former. Curves are to guide the eye.}
\end{figure*}

The transient reflectivity signals of the GaP/Si sample also exhibit
an apparent periodic modulation, as extracted in Fig.~\ref{CFT}(a)
after subtracting a multi-exponential baseline. This oscillation is
not seen for bulk Si (Fig.~S2(b) in SM) or bulk GaP (not shown)
measured under the same conditions. Indeed, the frequency of this
oscillation, 2\,THz, has no counterpart in the previously reported
first- and second-order Raman spectra of bulk GaP or Si
\cite{Hoff1973, Wang1973}. Hereafter we refer to the 2-THz
oscillation as ``low-frequency mode (LFM)”. In the fast
Fourier-transform (FFT) spectra in Fig.~\ref{CFT}(b), we also see a
small overtone of the LFM at $\sim$4\,THz for
$\lambda_\textrm{pump}<$800\,nm.

The oscillatory reflectivity in Fig.~\ref{CFT}(a) can be fitted
reasonably to a damped harmonic oscillation:
\begin{equation}
f(t)=A_\textrm{LFM} \exp(-\Gamma_\textrm{LFM} t)\sin(2\pi\nu_\textrm{LFM} t+\phi_\textrm{LFM}).
\end{equation}
The fitting parameters $A_\textrm{LFM}, \Gamma_\textrm{LFM},
\nu_\textrm{LFM}$ and $\phi_\textrm{LFM}$ are plotted as a function
of $\lambda_\textrm{pump}$ in Figs.~\ref{CFT}(c)-(e). The amplitude
$A_\textrm{LFM}$ in Fig.~\ref{CFT}(c) exhibits a distinct resonant
peak at $\lambda_\textrm{pump}\simeq$900\,nm (photon energy of
1.4\,eV). This resonance behavior coincides with that of $\Delta
R_\textrm{max}$ in Fig.~\ref{TDGaPSi_Si}(b), indicating a strong
coupling of the LFM with the interface electronic state. We
therefore assign the LFM as an interface phonon mode coupled with
the electronic transition at the GaP/Si interface. We note that the
frequency $\nu_\textrm{LFM}$ [Fig.~\ref{CFT}(d)] exhibits an
apparent jump, whereas the initial phase $\phi_\textrm{LFM}$
[Fig.~\ref{CFT}(e)] shows a shift by $\pi$, both, at
$\lambda_\textrm{pump}\sim$800 nm, which corresponds to our probe
wavelength.

Phonons at semiconductor heterointerfaces have been studied
extensively by means of resonant Raman scattering spectroscopy on
GaAs/AlAs superlattices \cite{Meynadier1987, Gridin1988,
Popovic1989a, Popovic1989b, Mowbray1991, Spitzer1992, Zhang1996}.
These studies reported higher-order (multiple) scattering in
addition to the first-order scattering by a single phonon. The
intensities of the higher-order Raman scattering were enhanced
significantly by tuning the incident photon energy to, for example,
the transition between the sub-bands of a quantum well. Multiple
scattering appeared not only by phonons within the same crystal, but
also by those from two different materials, e.g. GaAs and AlAs.
Whereas most of these studies focused on sum-combination modes
appearing at high frequencies, there was also a report on a Raman
peak appearing at low frequency of 108 cm$^{-1}$ (3.2 THz)
\cite{Spitzer1992}. Based on its resonance behavior it was
attributed to a difference-combination mode involving GaAs and AlAs
optical phonons.

\begin{figure}
\includegraphics[width=0.425\textwidth]{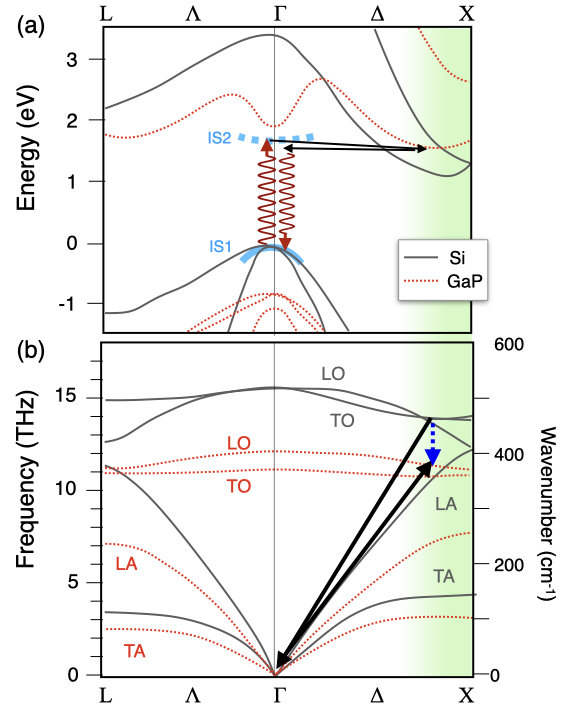}
\caption{\label{Band} (a) Schematic illustration of
doubly resonant second-order Raman scattering at the GaP/Si interface (arrows).
Calculated electronic structure of GaP \cite{Sieh1985} (orange curves) and
Si \cite{Ishioka2015} (gray curves) are superposed with a valence band
offset of 0.8 eV. IS1 and IS2 indicate occupied and unoccupied interface states
near the $\Gamma$-point, respectively. (b) Phonon dispersion curves of Si and GaP taken from Refs.~\cite{Weber1977, Borcherds1979}
with an example of a pair of Si and GaP optical phonons participating in the second-order scattering (arrows).}
\end{figure}

In the present study we similarly attribute the observed LFM to a
difference-combination mode between GaP and Si optical phonons. The
optical phonons of GaP and Si, both, reveal comparatively small
dispersion along the $\Delta-X$ direction of the Brillouin zone as
shown in Fig.~\ref{Band}b.  Thus, there is relatively high
density-of-states for a pair of Si and GaP optical phonons to
satisfy energy and momentum conservation, i.e.,
$\nu_\text{Si}-\nu_\text{GaP}=2$ THz and
$\bm{k}_\text{Si}=-\bm{k}_\text{GaP}$. An exemplary second-order
Raman scattering process that would give rise to the LFM is shown
schematically in Fig.~\ref{Band}(b). It is initiated by the creation
of electron-hole pairs upon optical excitation at the
$\Gamma$-point, followed by the scattering of a Si and a GaP optical
phonon and finally terminated by the charge-carrier recombination.

In Fig.~\ref{CFT}(d) and (e) we observed a small but clear
discontinuity in the LFM frequency and a phase shift by $\pi$ at
$\lambda_\textrm{pump}\simeq\lambda_\textrm{probe}$. The phase shift
is an indication for the Stokes and anti-Stokes processes in the
short ($\lambda_\textrm{probe}<\lambda_\textrm{pump}$) and long
($\lambda_\textrm{probe}>\lambda_\textrm{pump}$) probe wavelength
regimes, respectively \cite{Mizoguchi2013}. In principle, Stokes and
anti-Stokes scattering give rise to Raman peaks at the same
frequency for first-order scattering by a zone-center phonon. But
this is not necessarily the case for Raman scattering by
zone-boundary phonons combined with double (or multiple) resonance.
The frequency disparity between Stokes and anti-Stokes scattering
was most famously demonstrated in the disorder-induced Raman bands
($D, D', D"$ and their combination modes) of graphitic materials
\cite{Tan1998, Tan2002, Cancado2002, Zolyomi2002}. Correspondingly,
a time-resolved study on graphene also reported a drastic frequency
jump of the coherent $D$ band at
$\lambda_\textrm{probe}=\lambda_\textrm{pump}$ ~\cite{Katayama2013}.
The frequency jump observed in the present study also occurs at
$\lambda_\textrm{probe}\simeq\lambda_\textrm{pump}$, though not as
drastic, and therefore can similarly be interpreted as a double
resonance. Fig.~\ref{Band}(a) illustrates an example of a
second-order Stokes scattering with incoming resonance.  Here the
unoccupied interface state (IS2) and the GaP conduction band near
the $X$-point contribute as the real intermediate states and thereby
make the second-order scattering doubly resonant.  The existence of
such a double resonance can also explain the extraordinary large
amplitude of the LFM near the electronic resonance at
$E_\text{pump}=1.4$~eV.


In conclusion, our two-color transient reflectivity measurements
presented further experimental evidence for interface electronic
states at the buried GaP/Si(001) heterointerface.  In particular, an
interface phonon mode was unambiguously resolved as a periodic
modulation at $\sim$2 THz, whose amplitude was resonantly enhanced
by an electronic transition between the interface states. The
oscillation was interpreted as a difference second-order Raman
scattering involving a GaP and a Si optical phonon.  The unusual
photon-energy dependence of its frequency indicated the involvement
of a multiple resonance. We thus demonstrated electron-phonon
coupling that is characteristic to a heterointerface of polar and
non-polar inorganic semiconductors, whose knowledge is indispensable
in the construction of Si-based III/V optoelectronics.
		
\begin{acknowledgments}
We gratefully acknowledge funding  by the Deutsche
Forschungsgemeinschaft (DFG, German Research Foundation), Project-ID
223848855-SFB 1083. The authors thank NIMS RCAMC and NAsP III/V GmbH
for AFM measurements.
\end{acknowledgments}

\bibliographystyle{apsrev4-2}
\bibliography{GaPSiNOPA_ref}

\end{document}


\title{Supplementary Material for\\
Strongly coupled interface electronic state and interface phonon mode at GaP/Si(001)}

\author{Gerson Mette}
\affiliation{Faculty of Physics and Materials Sciences Center, Philipps-Universit{\"a}t Marburg, 35032 Marburg, Germany}

\author{Kunie Ishioka}
\affiliation{Nano-characterization Unit, National Institute for Materials Science, Tsukuba, 305-0047 Japan}

\author{Steven Youngkin}
\affiliation{Faculty of Physics and Materials Sciences Center, Philipps-Universit{\"a}t Marburg, 35032 Marburg, Germany}


\author{Christopher J. Stanton}
\affiliation{Department of Physics, University of Florida, Gainesville, FL 32611 USA}

\author{Wolfgang Stolz}
\affiliation{Faculty of Physics and Materials Sciences Center, Philipps-Universit{\"a}t Marburg, 35032 Marburg, Germany}

\author{Kerstin Volz}
\affiliation{Faculty of Physics and Materials Sciences Center, Philipps-Universit{\"a}t Marburg, 35032 Marburg, Germany}

\author{Ulrich H{\"o}fer}
\affiliation{Faculty of Physics and Materials Sciences Center, Philipps-Universit{\"a}t Marburg, 35032 Marburg, Germany}

\date{\today}

\maketitle

\section{Sample Fabrication and Characterization}

\begin{figure}[h]
\includegraphics[width=0.55\textwidth]{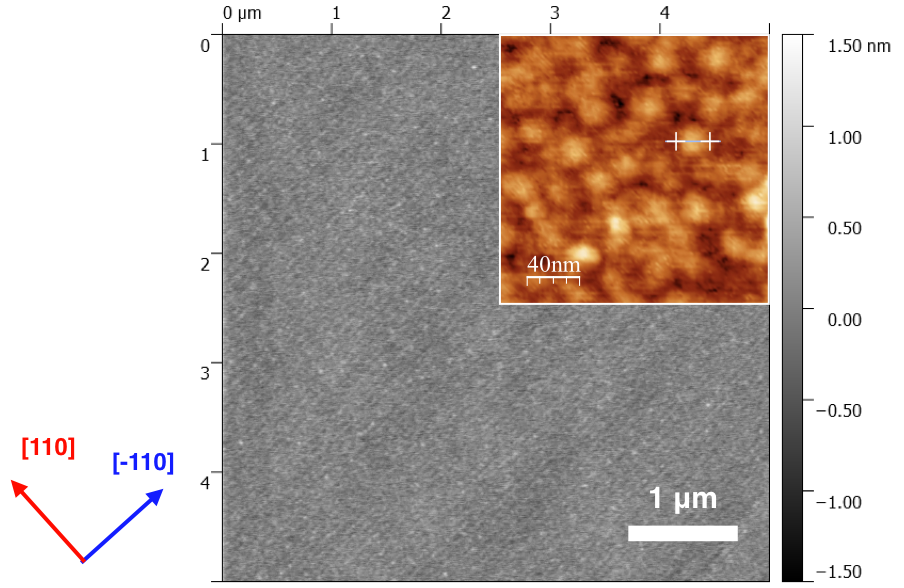}
\caption{\label{AFM} Atomic force microscopy image of the 10-nm thick GaP nucleation layer on Si(001).
}
\end{figure}

The studied GaP/Si heterostructure sample consists of a nominally
undoped GaP film with a nominal thickness of $d$=10\,nm which is
grown on an exact Si(001) substrate by metal-organic vapor phase
epitaxy (MOVPE)~\cite{Volz2011, Beyer2012}. First, a homoepitaxial
Si-buffer layer is deposited on an n-type ($\rho=0.007-0.02
\Omega$\,cm) Si(001) substrate with a 0.23$^\circ$ intentional
miscut in [110]-direction, and annealed in H$_2$ atmosphere to
stabilize double steps. A thin GaP nucleation layer is then grown on
the Si substrate by flow-rate modulated epitaxy (FME) at
450$^\circ$C, where the Ga and P precursors are injected
intermittently.  The as-grown GaP nucleation layer consists of
crystalline grains with a lateral size of $\leq$30\,nm as shown in
Fig.~\ref{AFM}.

\section{Spectroscopic experimental methods}

Two-color pump-probe reflectivity measurements are performed under
ambient conditions using two different sets of light sources to
cover the pump wavelength range of $720-1220$\,nm (photon energies
of $1.02-1.72$\,eV).
On the one hand, the output of a non-collinear optical parametric
amplifier (Orpheus-N-2H, LightConversion) with tuneable wavelength
range between 720 and 920\,nm (photon energies of $1.35-1.72$\,eV),
30-fs pulse duration and 200-kHz repetition rate is mainly used as
pump beam, whereas the output of an Orpheus-F-TWIN (LightConversion)
at 800\,nm (1.55\,eV photon energy), 40-fs duration and 200-kHz
repetition rate is used as probe beam.
On the other hand, to cover the longer pump wavelength, the idler
output of a collinear optical parametric amplifier (OPA~9450,
Coherent) with tuneable wavelength range between 845 and 1220\,nm
(photon energies of $1.02-1.47$\,eV ), 50-fs pulse duration and
100-kHz repetition rate is used as pump beam, whereas the output of
a regenerative amplifier (RegA 9050, Coherent) with 800-nm center
wavelength, 50-fs duration and 100-kHz repetition rate is used as
probe beam.
The photon energies employed in the present study are lower than the
indirect bandgap of GaP (2.26\,eV) or the direct gap of Si (3.4\,eV)
but comparable with or larger than the indirect gap of Si
(1.12\,eV).

The linearly polarized pump and probe beams are focussed to a
$\sim$30-$\mu$m diameter spot on the sample surface in the near
back-reflection geometry, with incident angles of $\lesssim15^\circ$
and $\lesssim5^\circ$ from the surface normal. Incident pump and
probe densities correspond to 4 and 0.7\,mJ/cm$^2$. The pump-induced
change $\Delta R$ in the reflectivity $R$ is measured by detecting
the probe lights before and after the reflection by the sample
(isotropic detection) with a pair of matched Si PIN photodiodes. The
pump beam is chopped at around 2\,kHz for lock-in detection. The
time delay $t$ between the pump and probe pulses is scanned using a
linear motor stage with a slow scan technique.


\section{Subtraction of the Si contribution from the as-measured transient reflectivity}

\begin{figure}[h]
\includegraphics[width=0.88\textwidth]{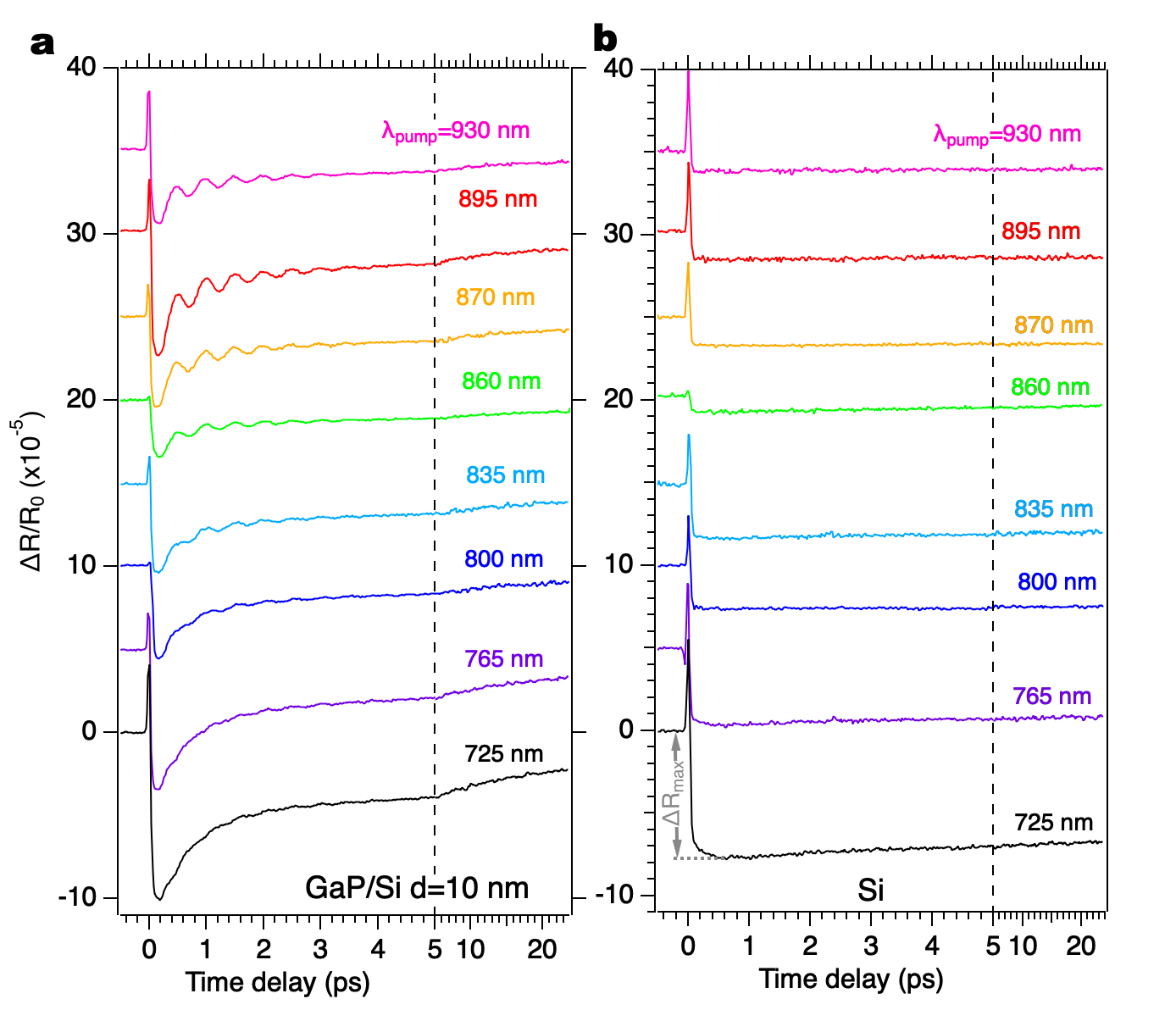}
\caption{\label{TDGaPSi}
As-measured transient reflectivity signals of the GaP/Si sample (a)
and of Si without a GaP overlayer (b) pumped at different wavelengths
and probed at 800\,nm.  The arrow in (b) indicates the initial step
height $\Delta
R_\textrm{max}$ for Si. Traces are offset for clarity.}
\end{figure}

The as-measured transient reflectivity signals from the GaP/Si
sample, plotted in Fig.~\ref{TDGaPSi}a, contain contributions from
the GaP overlayer and the heterointerface ($\Delta R_\text{int}$) as
well as from the Si substrate ($\Delta R_\text{sub}$).  We extract
the former contribution by subtracting the latter from the
as-measured signal $\Delta R$ as follows:
\begin{equation}\label{Si_sub2}
\dfrac{\Delta R_\text{int}(t)}{R_0}\equiv\dfrac{\Delta R(t)}{R_0}-\dfrac{\Delta R_\text{sub}(t)}{R_0}.
\end{equation}
The resulting transients are shown in Fig.~1a in the main text for
different pump wavelengths.

In Eq.~(\ref{Si_sub2}) the contribution from the Si substrate
depends on the overlayer thickness $d$, the pump wavelength
$\lambda_{pump}$ and the probe wavelength $\lambda_\text{probe}$ due
to interferences at the heterointerface. As discussed in detail in
Ref.~\cite{Ishioka2021}, it can be approximately given by:
\begin{equation}\label{Si_sub1}
\dfrac{\Delta R_\text{sub}(t,d)}{R_0}\equiv\dfrac{T_0(d,\lambda_\text{pump})P_2(d,\lambda_\text{probe})}{T_0(0,\lambda_\text{pump})P_2(0,\lambda_\text{probe})}\dfrac{\Delta R_\text{Si}(t)}{R_0},
\end{equation}
using the transient reflectivity of Si without GaP overlayer
measured at the same condition, $\Delta R_\text{Si}(t)$, which is
shown in Fig.~\ref{TDGaPSi}b.
$P_2(d,\lambda_\text{probe})$ in Eq.~(\ref{Si_sub1}) represents the
change in the reflected probe light intensity due to a pump-induced
disturbance in the refractive index $n_2$ of the Si substrate:
\begin{eqnarray}\label{omega2}
P_2(d, \lambda_\text{probe})&\equiv&
\dfrac{1}{R_0(d, \lambda_\text{probe})}\dfrac{\partial R_0}{\partial r_{12}}\dfrac{\partial r_{12}}{\partial n_2}\nonumber\\
&=&\dfrac{2r_{12}(1-r_{01}^4)+2r_{01}(1+r_{12}^2)(1-r_{01}^2)\cos2(n_1k_0d)}{(r_{01}^2+r_{12}^2+2r_{01}r_{12}\cos2n_1k_0d)(1+r_{01}^2r_{12}^2+2r_{01}r_{12}\cos2n_1k_0d)}\times\dfrac{-2n_1}{(n_1+n_2)^2}.
\end{eqnarray}
where $R_0$ denotes the reflectance of the heterointerface at
wavelength $\lambda$ and wavevector $k_0=2\pi/\lambda$ in air:
\begin{equation}\label{th20}
R_0(d, \lambda)=\left|\dfrac{r_{01}+r_{12}e^{4i\pi n_1k_0d/\lambda}}{1+r_{01}r_{12}e^{4i\pi n_1k_0d/\lambda}}\right|^2.
\end{equation}
$r_{01}$ and $r_{12}$ are the reflection coefficients for the light
waves incoming from air into GaP and from GaP into Si:
\begin{equation}
r_{01}=\dfrac{1-n_1}{1+n_1}; \quad r_{12}=\dfrac{n_1-n_2}{n_1+n_2},
\end{equation}
with $n_1$ and $n_2$ being the refractive indices of GaP and Si.
$T_0$  in eq.~(\ref{Si_sub1}) represents the transmittance of the
heterointerface, i.e., the pump intensity penetrating into the Si
substrate:
\begin{equation}
T_0(d, \lambda_\text{pump})=1-R_0(d, \lambda_\text{pump})=\dfrac{(1-r_{01}^2)(1-r_{12}^2)}{1+r_{01}^2r_{12}^2+2r_{01}r_{12}\cos{2n_1k_0d}}.
\end{equation}
%

\bibliographystyle{apsrev}
\bibliography{GaPSiNOPA_ref}